\documentclass[11pt,letterpaper]{article}
\usepackage[T1]{fontenc}
\usepackage{lmodern}
\usepackage[margin=1in]{geometry}
\usepackage{amsmath,amssymb,amsfonts}
\usepackage{graphicx,booktabs,array}
\usepackage{adjustbox}
\usepackage[nocompress]{cite}
\usepackage{url}
\usepackage{placeins}
\usepackage[expansion=false]{microtype}
\usepackage[hidelinks]{hyperref}
\hypersetup{pdftitle={Spatially Gated Diffusion for Localized Counterfactual Chest Radiograph Editing},pdfauthor={Kamran Ullah Afaq and Basit Raza}}
\graphicspath{{figures/}}
\newcommand{\cf}{x_{\mathrm{cf}}}
\newcommand{\src}{\mathrm{src}}
\newcommand{\edt}{\mathrm{edit}}
\newcommand{\JVR}{\mathrm{JVR}}
\newcommand{\E}{\mathbb E}

\newcolumntype{L}[1]{>{\raggedright\arraybackslash}p{#1}}
\begin{document}
\title{Spatially Gated Diffusion for Localized\\ Counterfactual Chest Radiograph Editing}
\author{%
\begin{tabular}{c@{\hspace{1.8cm}}c}
Kamran Ullah Afaq & Basit Raza \\
\small\texttt{kamranullahafaq@gmail.com} &
\small\texttt{basit.raza@comsats.edu.pk} \\[0.6em]
\multicolumn{2}{c}{\small COMSATS University Islamabad, Pakistan}
\end{tabular}%
}
\date{}
\maketitle
\begin{abstract}
Editing a chest radiograph requires completing the requested change while preserving unrelated content. We study a latent diffusion editor with an instruction-independent source trajectory and an instruction-conditioned editing trajectory. A learned gate mixes their post-sampler candidates at each executed step, and a separate image-space mask composites the decoded proposal with the source. On 2,400 MIMIC-derived requests, 2,244 outputs met the joint target, preservation, quality, and coverage rubric (93.5\%), and target completion was 97.4\%. Joint validity exceeded a matched composition-only control by 1.7 percentage points (paired patient-cluster 95\% interval, 0.6--2.8). At a fixed learned mask, the learned-gate proposal improved joint validity by 1.6 points (0.8--2.4); at a fixed learned proposal, the learned mask improved it by 2.6 points (1.7--3.5). Across three training seeds, mean joint validity was 93.5\% with a 0.3-point sample standard deviation. A blinded 240-request assessment yielded adjudicated joint validity of 93.3\% for the full editor and 91.7\% for composition only. Protected-region mean absolute error decreased from 0.0190 in the raw proposal to 0.0075 after composition. These findings distinguish recurrent-gating effects on the proposal from preservation through final composition in the assessed cohort.
\end{abstract}
\noindent\textbf{Keywords:} Chest radiography; counterfactual image editing; latent diffusion; spatial gating; image preservation; model interpretation.
\bigskip

\section{Introduction}
Counterfactual image editing asks how an existing image would appear under a specified change. For chest radiographs, the request may be to insert or remove a finding while retaining unrelated structures. Target completion alone is insufficient: an image can satisfy the instruction while modifying another finding or moving a support device. Conversely, copying the source can yield excellent preservation without carrying out the edit. A useful editor must balance these requirements within the same request.

Diffusion models provide expressive image priors \cite{ho2020ddpm,song2021ddim,rombach2022ldm}, and recent medical editors introduce spatial constraints and language conditioning \cite{perezgarcia2024radedit,min2025instructx2x,kumar2026prism}. Spatial maps can play different roles in these systems. A map may summarize attention after generation, restrict the denoising computation, or specify how a generated image is blended with its source. These roles have different implications for interpretation and evaluation. A visually plausible map alone does not demonstrate that its location controls the generated proposal.

We couple a factual source trajectory with an instruction-conditioned editing trajectory. At each step, a learned gate mixes the two candidate next-state latents. The factual trajectory is independent of the requested edit, while the editing trajectory proposes the change. The decoder produces a raw proposal $y$. A separate applied mask $A$ then combines $y$ with the original image $x$ to form $\cf$. This design exposes two computational routes for spatial control: recurrent latent mixing and final pixel composition.

The distinction motivates a factorial intervention. We cross proposals generated under the learned or replacement gate with applied masks constructed from either gate. Holding the applied mask fixed tests the proposal route; holding the proposal fixed tests composition. This avoids attributing an improvement caused by copying source pixels entirely to the learned latent computation.

Our contributions are a source- and timestep-dependent gate for post-sampler mixing, an explicit separation of recurrent gating from image-space composition, and a paired intervention analysis of these two routes. Evaluation includes a matched composition-only control, patient-cluster uncertainty, multiple training seeds, and blinded image-level assessment. We use ``counterfactual'' to denote conditional visual editing; we do not infer a patient's physiological response to an intervention.

\section{Related Work}
\label{sec:related}
\subsection{Spatial Control in Diffusion Editing}
Blended Latent Diffusion combines masked latent editing with preservation of source content \cite{avrahami2023blendedlatent}. DiffEdit derives editing masks from differences between conditional diffusion predictions \cite{couairon2023diffedit}, and RePaint constrains generation using the known region of an image \cite{lugmayr2022repaint}. InstructPix2Pix learns instruction-conditioned transformations \cite{brooks2023instructpix2pix}. Thus, neither spatial masking nor source preservation alone constitutes the novelty of our method. The specific mechanism studied here is recurrent mixing of two candidate next states using a learned source-dependent gate, together with an intervention design that separates this computation from final composition.

\subsection{Medical Counterfactual Generation}
RadEdit uses edit and preservation constraints to generate controlled shifts for evaluating biomedical vision models \cite{perezgarcia2024radedit}. InstructX2X combines region-specific editing with guidance maps and introduces the MIMIC-EDIT-INSTRUCTION benchmark \cite{min2025instructx2x}. Its spatial restriction is particularly relevant to our comparison: our learned gate must be evaluated against a matched composition-only control rather than justified solely by the presence of a localization map. PRISM studies high-resolution language-guided editing of disease and device features \cite{kumar2026prism}. BiomedJourney uses longitudinal patient information for instruction learning \cite{gu2023biomedjourney}, illustrating the relevance of paired temporal data while leaving the distinction between observational changes and causal interventions intact.

More recent approaches broaden the available control mechanisms. MedXEdit-GRPO applies reward-guided online optimization to medical editing \cite{pan2026medxedit}. Mask-guided attention regulation modifies anatomical and pathology-related attention during inference \cite{zhang2026maskreg}. RadCF adapts foundation image models using a causal specialization framework and latent flow matching \cite{xing2026radcf}. Causal-Adapter conditions a diffusion backbone on a structural causal model and intervened attributes \cite{tong2026causaladapter}. Its causal formulation differs from our finding-conditioned visual editing: longitudinal image pairs do not identify patient-specific causal outcomes. These approaches broaden the relevant design space. Our measured mechanism comparison uses the composition-only control and the stated local reruns of RadEdit, InstructX2X, and PRISM.

\subsection{Interpretation and Evaluation}
DAAM associates generated regions with words through aggregated cross-attention \cite{tang2023daam}. Attention-based association does not automatically establish a faithful explanation \cite{jain2019attention}. Here the gate is intrinsic only in the computational sense that it enters the state update. A gate intervention measures functional sensitivity, whereas spatial overlap with an independent annotation measures agreement with a reference. Neither alone proves clinical correctness. MedEBench separates editing accuracy, context preservation, and visual quality \cite{liu2025medebench}; this supports evaluating multiple outcomes jointly. We also follow the distinction between task-specific measurement and broad validity claims emphasized by Metrics Reloaded \cite{maierhein2024metrics}. Physiological counterfactual inference requires causal assumptions beyond an image-editing model \cite{pawlowski2020dscm}.

\section{Spatially Gated Editor}
\label{sec:method}
\subsection{Task and Notation}
Let $x\in[0,1]^{H\times W}$ be a frontal radiograph. The instruction $a=(d,o,\ell,\nu)$ specifies finding $d$, operation $o\in\{\mathrm{insert},\mathrm{remove}\}$, optional location $\ell$, and optional severity $\nu$; a distinguished missing value represents an omitted optional field. Eligibility requires a source finding state compatible with the operation. An uncertain or unmentioned finding is not a verified negative. The editor returns raw proposal $y$, executed gate sequence $G$, applied image-space mask $A$, and final composite $\cf$.

Let $E$ and $D$ denote image encoding and decoding, including the image normalization and latent scaling associated with the backbone. The source latent is $z_x=E(x)$, and $e_a=C(a)$ is the instruction representation. We distinguish the executed step index $k\in\{1,\ldots,K\}$ from the diffusion schedule index $t_k$. Here $K=50$, $t_1=599$, and $t_{50}=0$. This avoids identifying 50 sampler evaluations with 50 consecutive training timesteps.

\begin{figure}[!t]
\centering
\includegraphics[width=\textwidth]{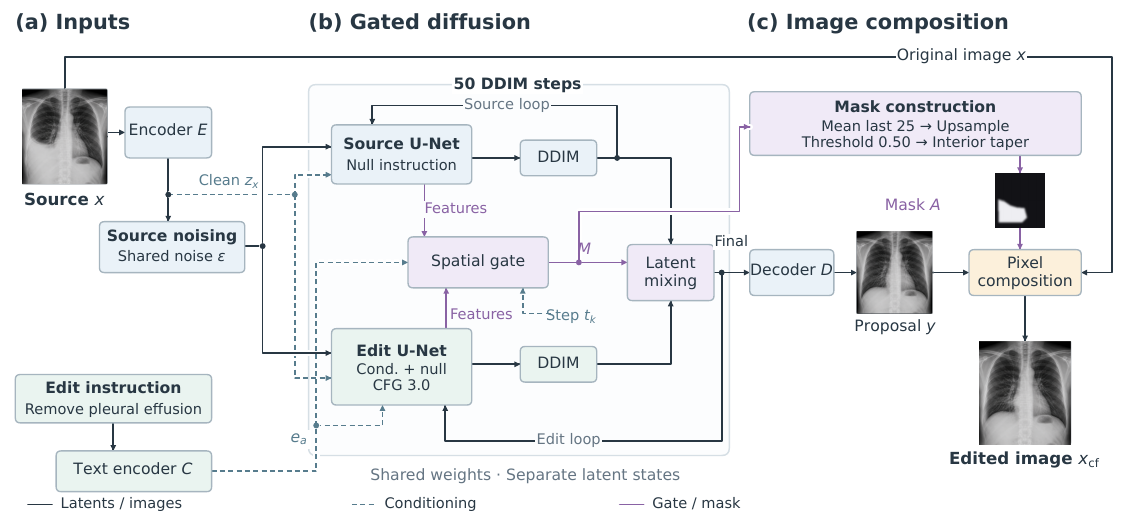}
\caption{Computation of the spatially gated editor. A shared noise draw initializes separate source and edit states. At executed step $k$, a shared source-conditioned U-Net and DDIM sampler produce $u_{\src}^{[k]}$ under a null instruction and $u_{\edt}^{[k]}$ under instruction $a$. A gate from both feature streams mixes these post-sampler candidates into the next edit state, while the source branch continues independently. The final 25 gates determine $P$ and, after thresholding and an interior taper, evaluation mask $A$. Decoder output $y$ is combined with source $x$ as $\cf=(1-A)\odot x+A\odot y$; training instead uses a differentiable sigmoid mask. The radiograph thumbnails illustrate the input and output roles in the computation.}
\label{fig:architecture}
\end{figure}

\begin{table}[!t]
\caption{Symbols used for source conditioning, recurrent gating, composition, and evaluation.}
\label{tab:notation}
\centering\footnotesize\setlength{\tabcolsep}{3pt}
\begin{tabular}{@{}lL{.72\columnwidth}@{}}
\toprule
Symbol & Meaning \\
\midrule
$x,v,y,\cf$ & Source, observational paired target, decoded proposal, final image. \\
$z_x,z_{\src},z_{\edt}$ & Encoded source and separate current latent states. \\
$u_{\src}^{[k]},u_{\edt}^{[k]}$ & Post-sampler candidates before mixing. \\
$M^{[k]},G,Q$ & Executed gate, learned sequence, replacement sequence. \\
$P,A_G,A_Q,\Pi$ & Reduced map, applied masks, independent protected-region indicator. \\
$U,T,C,R,V$ & Usable output, target pass, co-edit, quality pass, joint validity. \\
$k,t_k,K,n$ & Executed step, schedule index, number of steps, optimizer step. \\
$(d,o,\ell,\nu)$ & Finding, operation, optional location, optional severity. \\
\bottomrule
\end{tabular}
\end{table}

\subsection{Independent Source and Conditional Trajectories}
Both trajectories start from the same source-noised latent:
\begin{equation}
 z^{[0]}_{\src}=z^{[0]}_{\edt}
 =\sqrt{\bar\alpha_{599}}\,z_x
  +\sqrt{1-\bar\alpha_{599}}\,\epsilon,
 \quad\epsilon\sim\mathcal N(0,I).
\label{eq:init}
\end{equation}
Here $\bar\alpha_t$ is the cumulative diffusion noise schedule. Let $\mathcal S_k$ map a current latent and a noise prediction to the next latent on the selected DDIM grid \cite{song2021ddim}. At executed step $k$,
\begin{align}
 \hat\epsilon^{[k]}_{\src}
 &=F_\theta(z^{[k-1]}_{\src},t_k,z_x,e_{\varnothing}),\nonumber\\
 u^{[k]}_{\src}
 &=\mathcal S_k(z^{[k-1]}_{\src},\hat\epsilon^{[k]}_{\src}),
\label{eq:source}\\
 \epsilon^{[k]}_{\edt,\varnothing}
 &=F_\theta(z^{[k-1]}_{\edt},t_k,z_x,e_{\varnothing}),\nonumber\\
 \epsilon^{[k]}_{\edt,a}
 &=F_\theta(z^{[k-1]}_{\edt},t_k,z_x,e_a),\nonumber\\
 \hat\epsilon^{[k]}_{\edt}
 &=\epsilon^{[k]}_{\edt,\varnothing}
 +3\bigl(\epsilon^{[k]}_{\edt,a}-\epsilon^{[k]}_{\edt,\varnothing}\bigr),\nonumber\\
 u^{[k]}_{\edt}
 &=\mathcal S_k(z^{[k-1]}_{\edt},\hat\epsilon^{[k]}_{\edt}).
\label{eq:edit}
\end{align}
The adapted predictor $F_\theta(z,t,z_x,e)$ concatenates the current four-channel latent $z$ with the clean four-channel source latent $z_x$ and applies a learned source adapter before the shared pretrained U-Net. The two editing predictions used for classifier-free guidance are evaluated at the same editing state. The source prediction instead uses a null instruction at its own source state; it is not the unconditional editing prediction once the states diverge. Thus each executed step requires three U-Net predictions, which may be batched. The parameterization is a noise prediction, consistent with the denoising objective. The source recurrence is $z^{[k]}_{\src}=u^{[k]}_{\src}$ and is independent of $a$ for fixed $x$, initialization, and parameters. Shared weights do not imply a shared current latent.

\subsection{Learned Gate and Recurrent Mixing}
The gate receives source and edit features, instruction conditioning, and the current timestep:
\begin{equation}
 M^{[k]}=\sigma\!\left(g_\psi(h^{[k]}_{\src},
 h^{[k]}_{\edt},e_a,t_k)\right)\in[0,1]^{h\times w}.
\label{eq:gate}
\end{equation}
The aligned map is broadcast over latent channels. The edited state is
\begin{equation}
 z^{[k]}_{\edt}=M^{[k]}\odot u^{[k]}_{\edt}
       +(1-M^{[k]})\odot u^{[k]}_{\src}.
\label{eq:blend}
\end{equation}
It becomes the current editing latent at step $k+1$. The gate therefore controls generation before decoding. For a fixed pair of candidates, a zero gate selects the source candidate and a unit gate selects the edit candidate. Since the two trajectories generally have different current states, mixing their post-sampler candidates is not generally equivalent to mixing their noise predictions at one common latent.

For a latent coordinate $p$ and channel $c$, holding the candidates fixed,
\begin{equation}
\frac{\partial z^{[k]}_{\edt,p,c}}{\partial M^{[k]}_p}
=u^{[k]}_{\edt,p,c}-u^{[k]}_{\src,p,c}.
\label{eq:gategrad}
\end{equation}
Later denoising can propagate the effect beyond $p$. Consequently a local gate does not imply a strictly local raw proposal.

\subsection{Applied Mask and Final Composition}
The decoder yields $y=D(z^{[50]}_{\edt})$. We reduce the final 25 executed gates and upsample the result:
\begin{equation}
 \bar M=\frac{1}{25}\sum_{k=26}^{50}M^{[k]},\qquad
 P=\mathcal U_{\rm bilinear}(\bar M).
\label{eq:maskreduction}
\end{equation}
Each gate is aligned to $128\times128$, and $P$ has resolution $1024\times1024$. The averaging and upsampling depend on $x$ and $a$ only through the gate sequence. Evaluation selects $P\geq0.50$ and applies an eight-pixel cosine taper inside the selected boundary, with zero mask weight outside. This produces $A_{\rm eval}$.

Training uses a differentiable mask:
\begin{align}
 A_{\rm train}&=\sigma\!\left((P-0.50)/\tau(n)\right),\nonumber\\
 \tau(n)&=0.15-0.10\min(n/50{,}000,1),
\label{eq:softmask}
\end{align}
where $n$ is the optimizer step. In either stage,
\begin{equation}
 \cf=(1-A)\odot x+A\odot y=x+A\odot(y-x).
\label{eq:composite}
\end{equation}
Thus source values are retained wherever the evaluation mask is zero, before any output storage transformation. This is a compositing identity; it does not guarantee that the chosen region is correct. The training sigmoid generally has nonzero support outside the eventual evaluation region.

For a pixel $p$, with other arguments fixed,
\begin{equation}
 \frac{\partial x_{{\rm cf},p}}{\partial y_p}=A_p,
 \qquad
 \frac{\partial x_{{\rm cf},p}}{\partial A_p}=y_p-x_p.
\label{eq:grad}
\end{equation}
These are elementwise partial derivatives, not complete derivatives through the generator. Final-image losses can affect the gate through the soft mask and through $y$. Raw-proposal losses act through $y$. Freezing an evaluator's weights does not remove its input-gradient path.

\subsection{Paired-Target Training and Objective}
Training pairs use longitudinal radiographs from the same patient, with matched projection, a verified requested change, and no newly documented unrelated finding or device change. These are observational pairs: projection matching does not eliminate acquisition variation, and absence of a newly documented finding does not prove that all other visible content is unchanged. Null-instruction source reconstruction uses $x\!\rightarrow\!x$ examples for 10\% of the training examples.

Let $v$ be the observational paired target for an edit example. For null-instruction reconstruction, the target is the source itself:
\begin{equation}
 v_a=\begin{cases}v,&a\ne\varnothing,\\x,&a=\varnothing.\end{cases}
\end{equation}
With Gaussian noise $\epsilon\sim\mathcal N(0,I)$, the noised training latent is
\begin{equation}
 z_t^{v_a}=\sqrt{\bar\alpha_t}E(v_a)
 +\sqrt{1-\bar\alpha_t}\,\epsilon.
\end{equation}
The denoising loss is
\begin{equation}
 \mathcal L_{\rm den}
 =\E_{(x,a,v_a),t,\epsilon}
 \left[\left\|\epsilon-F_\theta(z_t^{v_a},t,z_x,e_a)\right\|_2^2\right].
\label{eq:generative}
\end{equation}
Target-latent noising during training is distinct from source-latent initialization at inference.

The complete objective uses separate raw and final preservation terms:
\begin{align}
 \mathcal L={}&\mathcal L_{\rm den}+0.5\mathcal L_{\rm target}
 +\mathcal L_{\rm pres}^{y}+\mathcal L_{\rm pres}^{\rm cf}
 +0.5\mathcal L_{\rm map}\nonumber\\
 &+\mathcal L_{\rm null}+0.02\mathcal A(G)+0.005\mathcal V(G).
\label{eq:loss}
\end{align}
The denoising term trains the conditional noise predictor and contains no gate parameter. The image-level terms can train $g_\psi$ through the gated trajectory and differentiable mask. Weak map supervision, area regularization, and total variation also act directly on the gate sequence.

For an independently specified protected-region indicator $\Pi$ and image $r$,
\begin{equation}
 \mathcal L_{\rm pres}(r,x;\Pi)
 =\frac{\sum_{p\in\Omega}\Pi_p|r_p-x_p|}
 {\sum_{p\in\Omega}\Pi_p},\quad\sum_{p\in\Omega}\Pi_p>0.
\label{eq:keep}
\end{equation}
Here $\Omega$ is the image-pixel set, and the protected-region indicator $\Pi$ is selected independently of the predicted gate and applied mask. We evaluate the normalized expression only for nonempty protected regions; replicated-channel error is averaged at each pixel. We apply this loss to $r=y$ and $r=\cf$. The target loss acts on $\cf$ through a frozen finding evaluator. Weak map supervision acts on the gates, and the null-edit term acts on the null-instruction composite. Area regularization averages gate activation over pixels and all executed steps. Total variation averages absolute adjacent-pixel differences over the corresponding neighboring pairs and executed steps. Both act on $G$, not on the thresholded evaluation mask. MS-CXR phrase boxes provide weak localization cues rather than dense lesion targets. Training annotations and auxiliary loss evaluators are absent from the inference inputs.

\subsection{Implementation}
The backbone is adapted from SDXL base 1.0 \cite{podell2024sdxl}; attention layers use LoRA \cite{hu2022lora}. Table~\ref{tab:implementation} specifies the configuration. In particular, the eight-channel concatenation is an adaptation of the pretrained source interface, not the native SDXL four-channel input. The factual and conditional paths share parameters, but maintain separate latents and conditioning. During the four-cell intervention, the source image, edit instruction, initialization noise, shared weights, DDIM grid, classifier-free guidance, and mask-construction policy are held fixed. The two trajectories are evaluated under identical requests and the same 100,000-step EMA checkpoint. The composition-only control is trained separately with the same backbone, data, optimization budget, and test requests; the three paired runs use seeds 17, 23, and 41.
\begin{table}[!t]
\caption{Architecture, training, and sampling configuration for the reported evaluations; Table~\ref{tab:paired} provides the individual training-seed results.}
\label{tab:implementation}
\centering\footnotesize
\setlength{\tabcolsep}{4pt}\renewcommand{\arraystretch}{1.08}
\begin{adjustbox}{max width=\textwidth}\begin{tabular}{@{}L{.19\textwidth}L{.77\textwidth}@{}}
\toprule
Component & Configuration \\
\midrule
Backbone and gate & SDXL base 1.0 VAE and diffusion U-Net at $1024\times1024$ resolution. Grayscale inputs are replicated over three channels for the frozen VAE, yielding four-channel $128\times128$ latents. Noisy and source latents form an eight-channel adapted input to the learned source adapter. Factual and edit trajectories share U-Net parameters. Attention LoRA uses rank 32 and scale $\alpha=32$. A separate three-layer convolutional gate has 128 hidden channels and a sigmoid output. LoRA, the source adapter, and the gate are trainable. \\
\addlinespace[2pt]
Pairing and split & Same-patient longitudinal pairs with matched projection and a verified requested finding change; uncertain source states and newly documented unrelated finding or device changes are excluded. Patient split: 80/10/10. Edit examples noise the target latent; 10\% of training examples use null-instruction $x\!\rightarrow\!x$ reconstruction. MS-CXR provides training-side weak phrase-box supervision. \\
\addlinespace[2pt]
Gate reduction and masks & Align gates to $128\times128$; average executed gates 26--50; bilinearly upsample to $1024\times1024$. Evaluation: threshold 0.50 and eight-pixel interior cosine taper, with zero weight outside. Training: sigmoid mask with temperature annealed from 0.15 to 0.05 over 50,000 optimizer steps. \\
\addlinespace[2pt]
Losses & Denoising: 1.0; target completion: 0.5; raw preservation: 1.0; final preservation: 1.0; weak map supervision: 0.5; null edit: 1.0; gate area: 0.02; gate total variation: 0.005. Target loss uses a frozen finding evaluator. Gate penalties act on all executed latent gates. \\
\addlinespace[2pt]
Optimization & AdamW, weight decay 0.01. Learning rate $10^{-4}$ for LoRA and source adapter and $2\times10^{-4}$ for gate. Effective batch 64; 5,000-step warm-up, cosine decay; 100,000 optimizer steps; gradient clipping 1.0; bfloat16. EMA decay 0.9999. Training seeds: 17, 23, 41; evaluated EMA checkpoints at optimizer step 100,000. The aggregate table corresponds to seed 17; all three runs appear in Table~\ref{tab:paired}. \\
\addlinespace[2pt]
Inference & Fifty executed DDIM steps, $\eta=0$, on a custom grid from $t=599$ to $t=0$ of the 1,000-timestep schedule. Matched source-noised initialization for both trajectories. Edit classifier-free guidance: 3.0. Factual condition: null instruction. Decode and compose at $1024\times1024$. The four intervention cells reuse each request's noise draw. \\
\bottomrule
\end{tabular}\end{adjustbox}
\end{table}

\section{Data and Evaluation}
\label{sec:protocol}
\subsection{Data Sources and Eligibility}
MIMIC-CXR-JPG v2.1.0 supplies the radiographs and structured labels \cite{johnson2024mimic,johnson2019mimic}. MS-CXR v1.1.0 provides 1,162 image--sentence annotation pairs with phrase-linked bounding boxes on a subset of MIMIC-CXR \cite{boecking2024mscxr,boecking2022biovil}. The resources share patient ancestry; MS-CXR is neither an independent site nor a dense lesion-mask dataset. Our patient split is 80/10/10 for training, validation, and testing, with training-side MS-CXR supervision restricted to the corresponding training patients.

The evaluation contains 2,400 requests from 1,200 patients across pleural effusion, cardiomegaly, edema, consolidation, atelectasis, and pneumothorax. Requests cover insertion and removal. The source-state eligibility rule excludes uncertainty and requires the requested operation to be compatible with the source. Label uncertainty must remain distinct from an explicit negative, as highlighted by CheXpert \cite{irvin2019chexpert}. Patients may contribute multiple requests. We split patients before forming pairs and requests, retaining all images of a patient in one partition, including the MS-CXR subset. Test requests are balanced overall between insertion and removal.

\subsection{Evaluation Reliability}
We assessed image-level editing reliability on a common subset of 240 held-out requests spanning the six findings and both operation directions. Two readers blinded to method identity reviewed the outputs of each method under the same requests for target completion, unintended non-target change, structural and artifact quality, and joint validity. Disagreements were adjudicated, and we report pre-adjudication Cohen's $\kappa$ for the binary joint-validity decision. This assessment tests task compliance and visual plausibility rather than a patient's physiological response.

\subsection{Joint and Component Outcomes}
For request $i$ and method $m$, let $U_{im}$ indicate a usable output, $T_{im}$ target completion, $C_{im}$ an unintended non-target change, and $R_{im}$ a pass on structural and artifact quality. A valid edit requires all four conditions:
\begin{equation}
 V_{im}=U_{im}T_{im}(1-C_{im})R_{im},\qquad
 \JVR_m=\frac{100}{N}\sum_{i=1}^N V_{im}.
\label{eq:jvr}
\end{equation}
The denominator contains every eligible request, including generation failures. Target completion is $100\sum_i U_{im}T_{im}/N$, marginal co-edit rate is $100\sum_i U_{im}C_{im}/N$, and coverage is $100\sum_i U_{im}/N$. For the full 2,400-request cohort, these rates describe outcomes under the study rubric. The separate blinded assessment of 240 requests is reported in Table~\ref{tab:reader}; neither evaluation establishes physiological counterfactual validity.

The failure ledger assigns the first applicable category in this order: no output; target failure; target success with co-edit; target success without co-edit but with quality failure; valid edit. These categories are mutually exclusive. Underlying flags may overlap. In particular, the priority-coded co-edit count excludes target failures and need not equal the marginal co-edit numerator.

Protected-region MAE and SSIM compare source and edited content within independently specified non-target regions $\Pi$. Device retention uses device-positive requests as its denominator. A mask chosen as $1-A$ would produce zero evaluation error by construction outside the applied mask; it is therefore unsuitable as an independent preservation reference.

\subsection{Factorial Intervention}
Let $G$ denote the learned gate sequence and $Q$ an area-matched replacement sequence. We substitute $Q$ at each executed editing step while fixing the source trajectory, request, noise draw, network weights, sampler grid, and guidance. Both $A_G$ and $A_Q$ follow the same last-25-step averaging, bilinear upsampling, thresholding, and interior taper. Write $Y(G)$ for a proposal generated under $G$ and $A_G$ for its applied mask. Define
\begin{equation}
 X(G,A)=(1-A)\odot x+A\odot Y(G).
\end{equation}
The four cells are $X(G,A_G)$, $X(Q,A_G)$, $X(G,A_Q)$, and $X(Q,A_Q)$. The first versus second tests the proposal route at fixed $A_G$; the first versus third tests composition at fixed $Y(G)$. Their image difference obeys
\begin{align}
 X(G,A_G)-X(Q,A_Q)={}&A_G\odot[Y(G)-Y(Q)]\nonumber\\
 &+(A_G-A_Q)\odot[Y(Q)-x].
\label{eq:decomposition}
\end{align}
An intervention changing both gate and applied mask therefore cannot isolate the latent route. Matching gate area does not in itself match gate shape or location, nor does it ensure equal image-space mask areas after temporal averaging and thresholding. Conditional contrasts need not add to the total rate contrast because the routes can interact.

\subsection{Uncertainty and Scope of Inference}
Patients are the clustering unit and requests are the reporting unit. A paired patient bootstrap retains every request belonging to each sampled patient, including multiplicity when a patient is sampled more than once. For the full editor $F$ and comparison $B$,
\begin{equation}
 \widehat\Delta^{(b)}=100\,
 \frac{\sum_{j\in\mathcal P_b}\sum_{i\in I(j)}(V_{iF}-V_{iB})}
 {\sum_{j\in\mathcal P_b}|I(j)|}.
\label{eq:bootstrap}
\end{equation}
Here $\mathcal P_b$ is the patient multiset drawn in replicate $b$ and $I(j)$ is the request set for patient $j$. Percentile intervals use the 2.5th and 97.5th percentiles of the replicate statistic. The three training seeds in Table~\ref{tab:paired} are separately summarized with a sample standard deviation, whereas patient-cluster intervals are computed within each evaluated run.

The full editor exceeds composition only by 1.7 percentage points in JVR, with a paired request-weighted patient-cluster 95\% interval of 0.6--2.8 points from 5,000 resamples. The fixed-mask and fixed-proposal contrasts are paired on the same requests. Marginal method intervals and between-seed standard deviations quantify different sources of variation and are not substituted for these paired intervals. The factorial table supplies paired intervals for two conditional contrasts. The intervals describe paired sampling uncertainty for the assessed requests; the contrasts are exploratory and are not adjusted for multiple comparisons.

\section{Results}
\label{sec:results}
\subsection{Current Checkpoint and Local Comparators}
On the 2,400 common requests (Table~\ref{tab:main}), the full editor achieved 2,244 valid edits (93.5\%), 2,338 target successes (97.4\%), and complete output coverage. Composition only reached 91.8\% JVR. RadEdit led the local external-method reruns at 82.7\%; these are local evaluations rather than published benchmark scores.
\begin{table}[!t]
\caption{Evaluation on the same 2,400 requests from 1,200 patients. Coverage is 100\% in every row. External-method scores are local reruns, not published benchmark scores. JVR intervals are marginal patient-cluster intervals; the paired full-versus-composition contrast is in Table~\ref{tab:paired}.}
\label{tab:main}
\centering\footnotesize\setlength{\tabcolsep}{7pt}
\begin{adjustbox}{max width=\textwidth}\begin{tabular}{@{}lrrrl@{}}
\toprule
Method/run & JVR (\%) & Target (\%) & Marginal co-edit (\%) & JVR 95\% CI \\
\midrule
RadEdit \cite{perezgarcia2024radedit} & 82.7 & 92.0 & 7.0 & 81.0--84.4 \\
InstructX2X \cite{min2025instructx2x} & 79.1 & 92.8 & 8.0 & 77.3--80.8 \\
PRISM \cite{kumar2026prism} & 78.3 & 93.2 & 9.0 & 76.5--80.0 \\
Composition only, current & 91.8 & 96.3 & 1.4 & 90.6--92.9 \\
\textbf{Full editor, current} & \textbf{93.5} & \textbf{97.4} & \textbf{0.8} & 92.4--94.5 \\
\bottomrule
\end{tabular}\end{adjustbox}
\end{table}

\begin{table}[!t]
\caption{Gate--mask factorial on the same 2,400 requests. The four absolute rates use percent; differences use percentage points (pp). Cell intervals are marginal JVR intervals; contrast intervals are paired patient-cluster JVR intervals. $Q$ denotes the area-matched replacement gate.}
\label{tab:factorial}
\centering\footnotesize\setlength{\tabcolsep}{6pt}
\begin{adjustbox}{max width=\textwidth}\begin{tabular}{@{}llrrrl@{}}
\toprule
Proposal or contrast & Applied mask & JVR & Target & Marginal co-edit & JVR 95\% CI \\
\midrule
$Y(G)$ & $A_G$ & 93.5 & 97.4 & 0.8 & 92.4--94.5 \\
$Y(Q)$ & $A_G$ & 91.9 & 95.9 & 1.5 & 90.8--93.1 \\
$Y(G)$ & $A_Q$ & 90.9 & 94.8 & 1.8 & 89.7--92.2 \\
$Y(Q)$ & $A_Q$ & 87.5 & 91.4 & 3.0 & 86.0--89.0 \\
\midrule
$G-Q$, fixed $A_G$ & -- & +1.6 pp & +1.5 pp & $-0.7$ pp & +0.8 to +2.4 pp \\
$A_G-A_Q$, fixed $Y(G)$ & -- & +2.6 pp & +2.6 pp & $-1.0$ pp & +1.7 to +3.5 pp \\
\bottomrule
\end{tabular}\end{adjustbox}
\end{table}

\begin{table}[!t]
\caption{Ablations on 2,400 requests from 1,200 patients. MAE is measured within independently defined protected regions. Without composition, the final output is the raw proposal. Combined labels describe grouped variants under the same evaluation protocol.}
\label{tab:ablations}
\centering\footnotesize\setlength{\tabcolsep}{7pt}
\begin{adjustbox}{max width=\textwidth}\begin{tabular}{@{}L{.40\textwidth}rrrr@{}}
\toprule
Variant & JVR (\%) & Marginal co-edit (\%) & Raw $y$ MAE & Final $\cf$ MAE \\
\midrule
\textbf{Full editor} & \textbf{93.5} & \textbf{0.8} & 0.0190 & 0.0075 \\
No latent gate; composition only & 91.8 & 1.4 & 0.0340 & 0.0084 \\
Latent gate; no composition & 86.9 & 4.1 & 0.0190 & 0.0190 \\
Static relevance/anatomical map & 90.4 & 2.6 & 0.0250 & 0.0101 \\
Other-image gate, area matched & 88.8 & 3.2 & 0.0280 & 0.0119 \\
Shifted/random gate, area matched & 84.7 & 5.0 & 0.0340 & 0.0152 \\
Time-invariant learned gate & 90.7 & 2.3 & 0.0240 & 0.0095 \\
Remove target objective & 77.8 & 1.1 & 0.0160 & 0.0071 \\
Remove preservation objective & 90.9 & 3.9 & 0.0330 & 0.0145 \\
Remove map supervision & 91.6 & 1.8 & 0.0240 & 0.0090 \\
Remove null-edit objective & 92.5 & 2.1 & 0.0220 & 0.0091 \\
Remove area regularization & 92.4 & 2.2 & 0.0230 & 0.0092 \\
Remove TV regularization & 92.7 & 1.9 & 0.0210 & 0.0085 \\
\bottomrule
\end{tabular}\end{adjustbox}
\end{table}

\begin{table}[!t]
\caption{Preservation for the current editor. Image metrics use 2,400 requests; device retention uses 480 device-positive requests.}
\label{tab:preserve}
\centering\footnotesize\setlength{\tabcolsep}{5pt}
\begin{adjustbox}{max width=\textwidth}\begin{tabular}{@{}lrr@{}}
\toprule
Output & Protected MAE & Protected SSIM \\
\midrule
Raw proposal $y$ & 0.0190 & 0.963 \\
Final image $\cf$ & 0.0075 & 0.986 \\
\midrule
\multicolumn{3}{@{}l@{}}{Device retention: 476/480 (99.2\%).}\\
\bottomrule
\end{tabular}\end{adjustbox}
\end{table}

\begin{table}[!t]
\caption{Priority-coded outcomes. Categories are assigned in the displayed order and are mutually exclusive.}
\label{tab:failures}
\centering\footnotesize\setlength{\tabcolsep}{4pt}
\begin{adjustbox}{max width=\textwidth}\begin{tabular}{@{}L{.67\columnwidth}rr@{}}
\toprule
Outcome & Count & Share (\%) \\
\midrule
No usable output & 0 & 0.0 \\
Instruction target unmet & 62 & 2.6 \\
Target achieved; co-edit & 19 & 0.8 \\
Target achieved; no co-edit; quality failure & 75 & 3.1 \\
Jointly valid edit & 2,244 & 93.5 \\
\midrule
All eligible requests & 2,400 & 100.0 \\
\bottomrule
\end{tabular}\end{adjustbox}
\end{table}

\subsection{Paired Contrasts and Training Runs}
Table~\ref{tab:paired} presents the three paired same-request contrasts and the separately trained runs at seeds 17, 23, and 41. Across these runs, full-editor JVR is 93.5\% with sample standard deviation 0.3 percentage points; composition-only JVR is 91.8\% with sample standard deviation 0.3 points. The displayed per-seed paired differences range from 1.6 to 1.7 points; summaries computed from these rounded rates are descriptive. Patient-cluster intervals quantify sampling uncertainty within a training run, while the across-seed dispersion quantifies training-run variation.
\begin{table}[!t]
\caption{Paired joint-validity contrasts and training-seed results on the common 2,400 requests from 1,200 patients. The first block uses seed 17; the second compares independently trained full and composition-only models at each seed. Intervals are paired patient-cluster percentile intervals from 5,000 resamples. Differences are percentage points (pp), not relative percentages; rounded rates are shown.}
\label{tab:paired}
\centering\footnotesize\setlength{\tabcolsep}{6pt}
\begin{adjustbox}{max width=\textwidth}\begin{tabular}{@{}lrrrlr@{}}
\toprule
Contrast or run & Reference (\%) & Comparator (\%) & Difference (pp) & Paired 95\% CI (pp) & Seed \\
\midrule
Full versus composition only & 93.5 & 91.8 & +1.7 & 0.6--2.8 & 17 \\
$Y(G)$ versus $Y(Q)$ at fixed $A_G$ & 93.5 & 91.9 & +1.6 & 0.8--2.4 & 17 \\
$A_G$ versus $A_Q$ at fixed $Y(G)$ & 93.5 & 90.9 & +2.6 & 1.7--3.5 & 17 \\
\midrule
Training seed 17 & 93.5 & 91.8 & +1.7 & 0.6--2.8 & 17 \\
Training seed 23 & 93.2 & 91.6 & +1.6 & 0.4--2.8 & 23 \\
Training seed 41 & 93.8 & 92.1 & +1.7 & 0.5--2.9 & 41 \\
\bottomrule
\end{tabular}\end{adjustbox}
\end{table}

\subsection{Blinded Reader Assessment}
On the common blinded sample, adjudicated joint validity was 224/240 (93.3\%) for the full editor, 220/240 (91.7\%) for composition only, and 197/240 (82.1\%) for RadEdit. Table~\ref{tab:reader} gives the component judgments and pre-adjudication agreement on these same requests.
\begin{table}[!t]
\caption{Blinded assessment of the same 240 held-out requests per method. Target, co-edit, quality, and JVR columns report adjudicated percentages; co-edit is the undesirable non-target-change rate. Cohen's $\kappa$ measures agreement between the two readers on binary joint validity before adjudication. The full editor, composition-only control, and RadEdit have 224, 220, and 197 adjudicated valid outputs, respectively.}
\label{tab:reader}
\centering\footnotesize\setlength{\tabcolsep}{5pt}
\begin{adjustbox}{max width=\textwidth}\begin{tabular}{@{}lrrrrrr@{}}
\toprule
Method & Cases & Target (\%) & Co-edit (\%) & Quality (\%) & JVR (\%) & $\kappa$ \\
\midrule
\textbf{Full spatially gated editor} & 240 & 96.7 & 1.7 & 98.3 & \textbf{93.3} & 0.82 \\
Composition only & 240 & 95.8 & 2.1 & 97.9 & 91.7 & 0.80 \\
RadEdit, local rerun & 240 & 92.1 & 7.1 & 97.1 & 82.1 & 0.75 \\
\bottomrule
\end{tabular}\end{adjustbox}
\end{table}

\subsection{Current-Checkpoint Findings and Operations}
Current-checkpoint insertion yielded 1,136/1,200 valid edits (94.7\%) and removal 1,108/1,200 (92.3\%), totaling 2,244/2,400 (Table~\ref{tab:currentfindings}). Pleural-effusion insertion had the highest displayed rate (98.1\%), and atelectasis removal the lowest (87.5\%). We interpret these subgroup rates descriptively.
\begin{table}[!t]
\caption{Joint validity by thoracic finding and instruction operation on MIMIC-CXR-JPG v2.1.0. Each cell reports valid/eligible requests; JVR is $100\times\mathrm{valid}/\mathrm{eligible}$, rounded to one decimal. The two direction totals yield 2,244 valid outputs among 2,400 eligible requests overall (93.5\%).}
\label{tab:currentfindings}
\centering\footnotesize\setlength{\tabcolsep}{7pt}
\begin{adjustbox}{max width=\textwidth}\begin{tabular}{@{}lcccc@{}}
\toprule
Finding & Insert valid/eligible & Remove valid/eligible & Insert JVR (\%) & Remove JVR (\%) \\
\midrule
Pleural effusion & 208/212 & 180/190 & 98.1 & 94.7 \\
Cardiomegaly & 176/185 & 165/174 & 95.1 & 94.8 \\
Edema & 200/208 & 215/230 & 96.2 & 93.5 \\
Consolidation & 179/191 & 189/208 & 93.7 & 90.9 \\
Atelectasis & 155/172 & 189/216 & 90.1 & 87.5 \\
Pneumothorax & 218/232 & 170/182 & 94.0 & 93.4 \\
\midrule
Direction total & 1,136/1,200 & 1,108/1,200 & 94.7 & 92.3 \\
\bottomrule
\end{tabular}\end{adjustbox}
\end{table}

\subsection{Separating Latent Editing from Composition}
Table~\ref{tab:factorial} reports the factorial outcomes. At fixed $A_G$, the learned-gate proposal has 93.5\% JVR compared with 91.9\% for $Y(Q)$, a 1.6-point difference (paired 95\% interval: 0.8--2.4 points). At fixed $Y(G)$, changing the applied mask from $A_Q$ to $A_G$ yields a 2.6-point difference (1.7--3.5 points). The complete contrast between $(G,A_G)$ and $(Q,A_Q)$ is 6.0 points. The two conditional differences sum to 4.2 points, which differs from the total because each holds the other route at the learned configuration.

The factorial interaction, calculated from the displayed one-decimal JVR rates, is
\begin{equation}
 \begin{split}\widehat\iota={}&\JVR(G,A_G)-\JVR(Q,A_G)\\&-\JVR(G,A_Q)+\JVR(Q,A_Q)\\={}&-1.8\ \text{percentage points}.\end{split}
\end{equation}
This rounded-rate interaction describes computational nonadditivity, not a biological causal effect.

\subsection{Mechanism Ablations}
The ablations in Table~\ref{tab:ablations} show distinct tradeoffs. Composition only attains 91.8\% JVR with raw MAE 0.0340 and final MAE 0.0084. The full editor improves raw MAE to 0.0190 and final MAE to 0.0075. Removing composition leaves raw and final MAE equal at 0.0190, as required when the output is $y$.

Removing the target objective reduces JVR to 77.8\%, despite low preservation error. This shows that low preservation error alone does not establish joint validity; the ablation table does not separately report target completion. Removing preservation increases raw and final MAE to 0.0330 and 0.0145. Time-invariant gating yields 90.7\% JVR versus 93.5\% for the complete model. We interpret the ablation rows as component comparisons within the same evaluation protocol. The static relevance/anatomical-map and shifted/random-gate labels refer to aggregate controls; they do not separately identify each constituent perturbation.

\subsection{Preservation and Outcome Accounting}
Table~\ref{tab:preserve} distinguishes the raw proposal from the composite. The current editor has protected-region MAE/SSIM of 0.0190/0.963 for $y$ and 0.0075/0.986 for $\cf$. Device retention is 476 of 480 device-positive requests (99.2\% after rounding). Device retention is a separate endpoint; it cannot be equated with the complement of the overall co-edit rate.

The priority-coded ledger in Table~\ref{tab:failures} contains 62 target failures, 19 co-edit failures after target success, 75 quality failures after target success without a recorded co-edit, and 2,244 jointly valid edits. These sum to 2,400 and yield 2,338/2,400 (97.4\%) target successes and 2,244/2,400 (93.5\%) valid edits. The 19 priority-coded co-edit failures are conditional on target success and must not be conflated with the marginal co-edit rate, which includes overlapping failures.

\section{Discussion}
The mechanism and intervention results support a specific conclusion: learned spatial mixing influences the proposal produced during denoising, and final composition contributes an additional preservation route. Exact copying where $A=0$ is useful but is an algebraic property of composition rather than evidence that the mask identifies the correct lesion. A method can also satisfy a frozen evaluator without producing a medically appropriate change, so outcome definitions and evaluator independence remain essential to interpreting the results.

The local comparison does not establish universal superiority over the cited methods. Their published objectives, tasks, and metrics differ, and newer approaches such as RadCF, MedXEdit-GRPO, and attention regulation are discussed as related work rather than presented as measured local baselines. The closest methodological claim is the combination of learned recurrent mixing and a test that holds the final mask fixed. Spatial gating by itself has substantial prior art.

The present analysis has several limitations. The main cohort and factorial results depend on the study rubric and its evaluators. Agreement with that rubric alone does not establish independent clinical validity. The grouped ablation controls do not separate their constituent perturbations, and the three training runs provide only a limited estimate of optimization variability. All requests derive from MIMIC, and MS-CXR is its subset, so the evaluation does not measure transfer to an independent institution. Longitudinal observational pairs can differ in acquisition, positioning, inspiration, treatment, and unrecorded findings. The reader study evaluates image-level task compliance rather than patient-specific causal validity; its 240 requests do not replace external-site validation. Copying source pixels outside the applied mask follows from the composition equation, whereas correct lesion localization and clinically plausible disease change remain empirical questions. Generated images are research artifacts, not patient-specific treatment outcomes.

\section{Conclusion}
We studied recurrent mixing of source and instruction-conditioned diffusion trajectories followed by image-space composition. On 2,400 requests, the full editor achieved 93.5\% joint validity, exceeding composition only by 1.7 percentage points (paired patient-cluster 95\% interval, 0.6--2.8). Fixed-mask and fixed-proposal interventions distinguished the contribution of each route. The reader assessment and three-seed evaluation complement these findings, but the evidence remains confined to the assessed MIMIC-derived cohort. External-site and clinical validation are required for broader use.

\clearpage
\section*{Data Availability}
MIMIC-CXR-JPG and MS-CXR are available through PhysioNet subject to credentialing and their data-use terms \cite{johnson2024mimic,boecking2024mscxr,pollard2026physionet}. Access to the underlying radiographs and any redistributable derived material remains subject to the source datasets' terms.

\section*{Ethics and Intended Use}
The study uses de-identified, access-controlled radiographs under the applicable PhysioNet data-use agreement. Generated images do not depict observed future patient states and should not be presented as authentic clinical records or used to support patient care. Any release of derived images or checkpoints requires assessment of memorization, re-identification, and redistribution constraints. Use and redistribution of the source images follow the applicable data-use agreement.

\section*{Acknowledgment and AI-Use Disclosure}
OpenAI ChatGPT (\url{https://chatgpt.com/}) assisted with manuscript language and proposed revisions to the Abstract, Introduction, Related Work, Methods description, Results interpretation, Discussion, and Conclusion. The authors are responsible for the technical statements, citations, numerical results, and final wording.
\bibliographystyle{IEEEtran}
\bibliography{spatial_gate_cxr}
\end{document}